\documentclass[aip,rsi,reprint]{revtex4-1}
\usepackage{braket}
\usepackage{amsmath}
\usepackage{dcolumn}
\usepackage{bm}
\usepackage{graphicx}
\begin{document}
\title{Active Stabilization of Ion Trap Radiofrequency Potentials}
\date{\today}
\author{K. G. Johnson}
\author{J. D. Wong-Campos}
\author{A. Restelli}	
\author{K. A. Landsman}
\author{B. Neyenhuis}
\author{J. Mizrahi}
\author{C. Monroe}
\affiliation{Joint Quantum Institute and University of Maryland Department of Physics, College Park, Maryland 20742, USA}
\begin{abstract}
We actively stabilize the harmonic oscillation frequency of a laser-cooled atomic ion confined in a rf Paul trap by sampling and rectifying the high voltage rf applied to the trap electrodes. We are able to stabilize the 1 MHz atomic oscillation frequency to better than 10 Hz, or 10 ppm.  This represents a suppression of ambient noise on the rf circuit by 34 dB. This technique could impact the sensitivity of ion trap mass spectrometry and the fidelity of quantum operations in ion trap quantum information applications.
\end{abstract}
\maketitle
\section{Introduction}
Charged particles are often controlled with radiofrequency (rf) electrical potentials, whose field gradients provide time-averaged (ponderomotive) forces that form the basis for applications such as quadrupole mass filters, ion mass spectrometers, and rf (Paul) ion traps \cite{Dehmelt, Paul}. These rf potentials, typically hundreds or thousands of volts at frequencies ranging from 1kHz to 100 MHz, drive high impedance loads in vacuum and are usually generated with rf amplifiers and resonant step-up transformers such as quarter-wave or helical resonators \cite{Siverns_12_APB}.  Such circuitry is susceptible to fluctuations in amplifier gain, mechanical vibrations of the transformer, and temperature drifts in the system.  Ion traps are particularly sensitive to these fluctuations, because the rf potential determines the harmonic oscillation frequency of the trapped ions.  Stable trap frequencies are crucial in applications ranging from quantum information processing \cite{WinelandBlatt08, MonroeKimScience} and quantum simulation \cite{Richerme2014, Jurcevic2014} to the preparation of quantum states of atomic motion \cite{Leibfried_03_RMP}, atom interferometry \cite{Johnson2015}, and quantum-limited metrology \cite{NISTClock2010}.
\begin{figure}
 	\includegraphics[scale=.3]{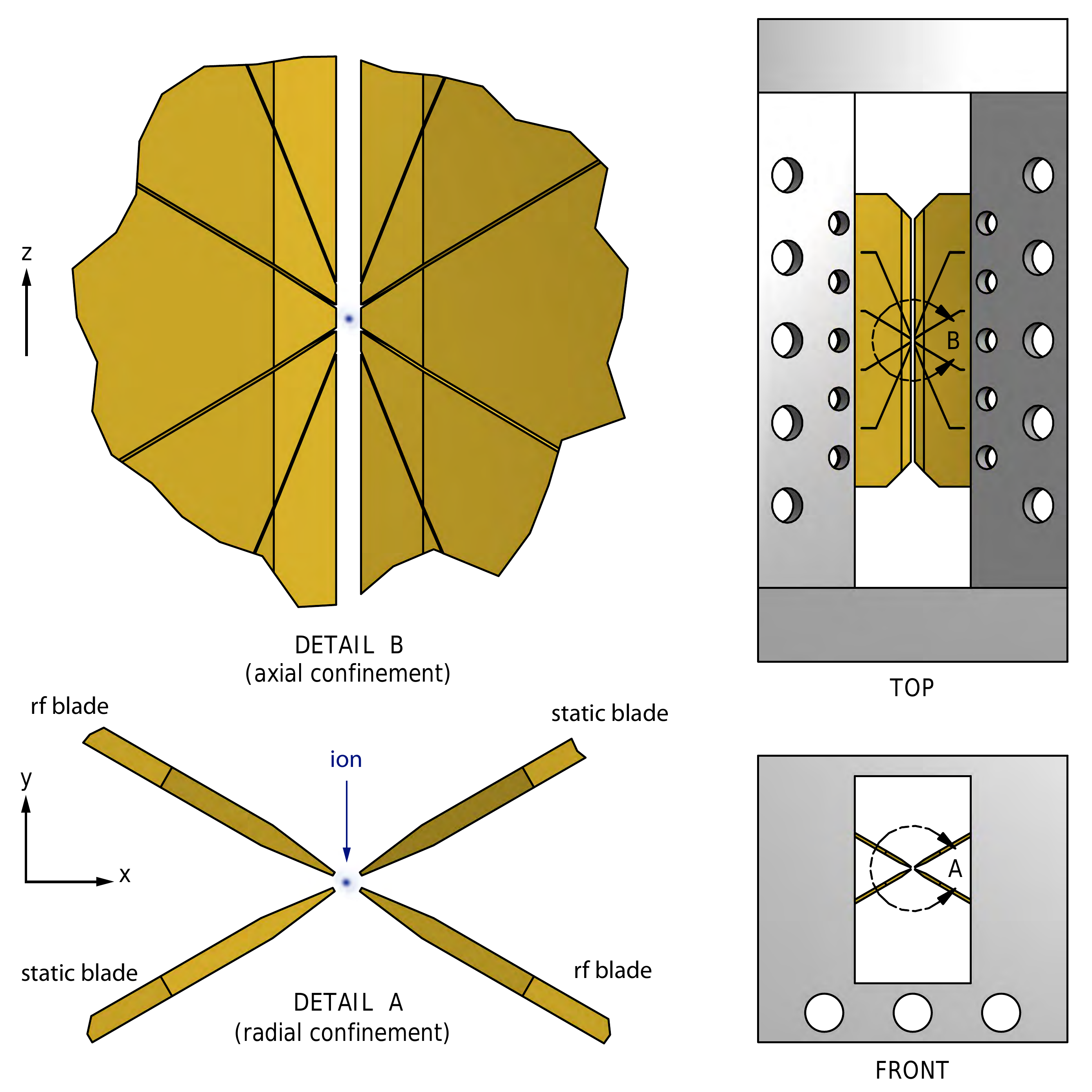}
 	\caption{\label{blade_trap_sections.pdf} Linear Paul trap created with four gold-plated blade electrodes that are held in place by an insulating mount.  An ion is confined in between the electrodes through a combination of rf and static potentials applied to the electrodes.  Each blade is split longitudinally into $5$ segments that are electrically isolated on the static blades and electrically connected on the rf blades.  The transverse distance from the ion axis to each electrode is $R=200$ $\mu$m, and the length of the central longitudinal segments is $400$ $\mu$m.}
\end{figure}
 \begin{figure*}
 	\includegraphics[scale=.35]{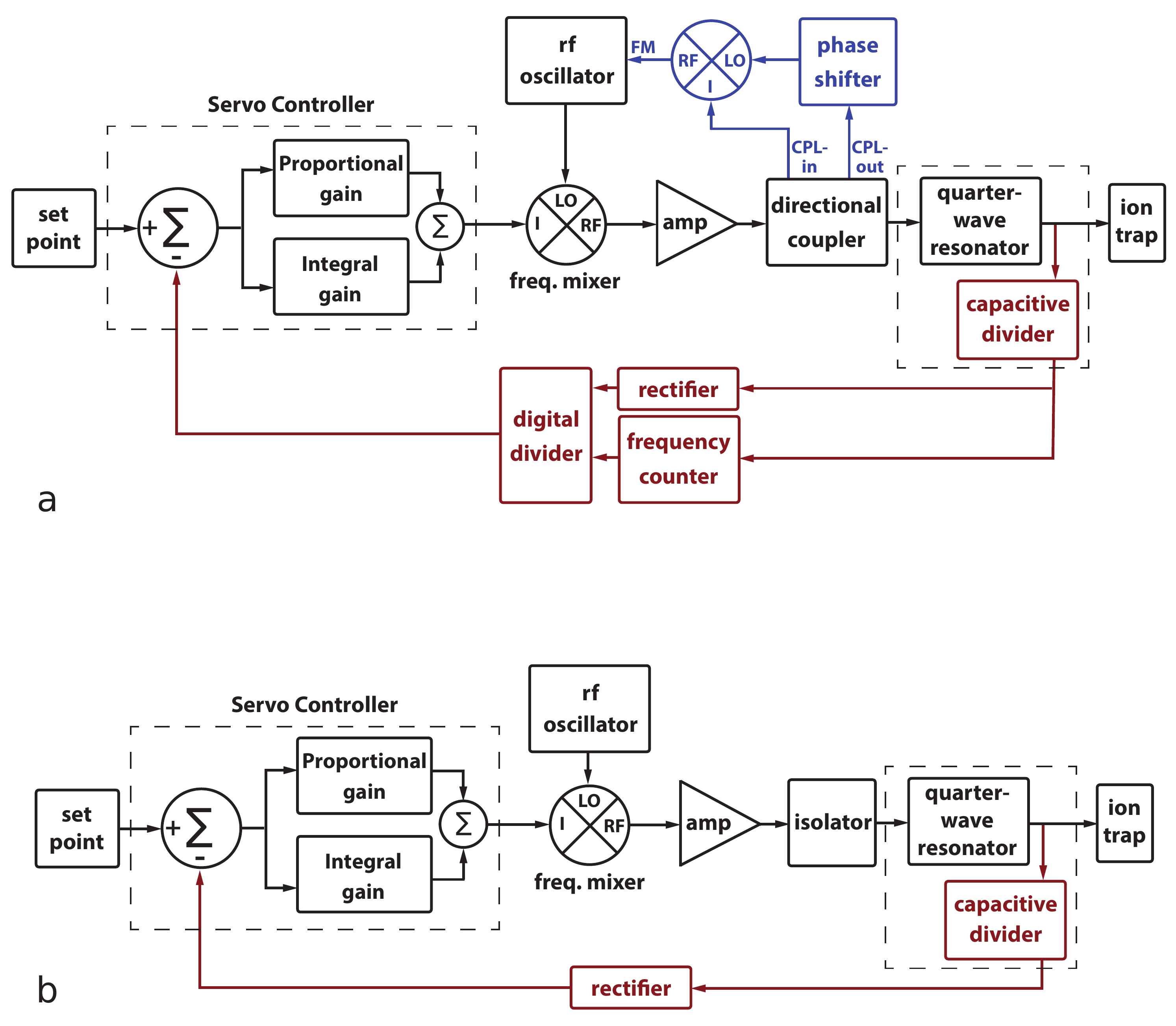}
 	\caption{\label{can_lock_schematic.pdf} Schematics of ion trap rf drive with active stabilization of the ion oscillation frequency $\omega$.  (a) Stabilization of the ratio of rf potential amplitude to frequency $V_0/\Omega$ (red), with a separate feedback loop (purple) that locks the rf drive frequency $\Omega$ to the resonant frequency of the step-up transformer. (b) Stabilization of the rf potential amplitude $V_0$ only, with fixed rf drive frequency (used in the experiment reported here).}
 \end{figure*}
 
Actively stabilizing rf ion trap potentials requires the faithful sampling of the rf potential.  Probing the signal directly at the electrodes is difficult in a vacuum environment and can load the circuit or spoil the resonator quality factor.  On the other hand, sampling the potential too far upstream is not necessarily accurate, owing to downstream inductance and capacitance.  Here we actively stabilize the oscillation frequency of a trapped ion by noninvasively sampling and rectifying the high voltage rf potential between the step-up transformer and the vacuum feedthrough leading to the ion trap electrodes.  We use this signal in a feedback loop to regulate the rf input amplitude to the circuit.  We stabilize a $1$ MHz trapped ion oscillation frequency to $< 10$ Hz after 200 s of integration, representing a 34 dB reduction in the level of trap frequency noise and drift, over a locking bandwidth of up to 30 kHz.

The ion is trapped in a linear rf trap, which consists of a two-dimensional rf quadrupole electric field superposed with a static quadrupole electric field to provide confinement along the longitudinal direction \cite{Raizen_1992_PRA}.  Longitudinal confinement is typically set much weaker than the transverse confinement, so that a crystal of laser-cooled ions can reside along the $x=y=0$ rf field null without feeling the effects of rf-induced micromotion \cite{Dehmelt}.  The transverse confinement, dictated by the rf fields, is used for many applications because motion along these directions is at higher frequency and the normal mode spectrum for a chain of ions can be tuned \cite{Zhu_2006_PRL}. Linear ion traps exist in a variety of topologically equivalent electrode configurations, even with electrodes all in a single plane for ease in lithographic fabrication \cite{Chiaverini_05_QIC}.  

The linear trap used in this experiment has four gold-plated ceramic ``blade" electrodes with their edges running parallel to the longitudinal (z) axis of the trap as shown in Fig. \ref{blade_trap_sections.pdf}. Two opposite blades are driven with an rf potential with respect to the other two static blades, creating the transverse (x-y) quadrupole confinement potential. Appropriate static potentials applied to the longitudinally-segmented static blades serve to confine the ions along the z-axis.  The rf electric quadrupole potential near the center of the trap $V(x,y)=\frac{\mu V_0}{2 R^2}(x^2-y^2)\cos\Omega t$ is set by the rf amplitude on the trap electrode $V_0$, the distance from the ion to the electrodes $R$, the rf drive frequency $\Omega$, and a dimensionless geometric efficiency factor $\mu\sim 0.3$ for the geometry of Fig. \ref{blade_trap_sections.pdf}. A particle with charge $e$ and mass $m$ inside the trap feels a resulting ponderomotive ``psuedopotential" $U_{\textrm{pon}}=\frac{e^2}{4m\Omega^2}|\nabla V|^2 = \frac{e^2 \mu^2 V_0^2}{4 m R^4 \Omega^2}(x^2+y^2)$, with harmonic oscillation frequency\cite{Paul},
\begin{equation}{\label{eq_secular_frequency}}
\omega=\frac{e \mu V_0}{\sqrt{2} m \Omega R^2}.
\end{equation}
This expression is valid under the pseudopotential approximation where $\omega \ll \Omega$ \cite{Paul, Dehmelt}, and we do not consider the residual transverse forces from the static potentials, because they are relatively small and stable.

One approach to stabilize the ion oscillation frequency is to control the ratio $V_0/\Omega$, in cases where the rf drive frequency is itself dithered to maintain resonance with the step-up transformer.  A feedback system of this style is shown in Fig. \ref{can_lock_schematic.pdf}a, where the lower feedback loop (red) stabilizes the ratio $V_0/\Omega$ and the upper feedback loop (purple) locks the variable rf oscillator frequency to the resonance of the transformer, which might drift due to mechanical or temperature fluctuations.  The main difficulty with this approach is the required performance of the digital divider circuit, which must have a precision as good as the desired stability, and be fast enough to stablize the system at at the desired bandwidth.  Moreover, higher order corrections to the trap frequency beyond the psuedopotental expression of Eq. \ref{eq_secular_frequency} depend on terms that do not scale simply as the ratio $V_0/\Omega$.  Therefore, we instead stabilize the rf potential amplitude $V_0$ alone, and use a fixed frequency rf oscillator and passively stable transformer circuit, as depicted in Fig. \ref{can_lock_schematic.pdf}b.

\begin{figure}
	\includegraphics[scale=.23]{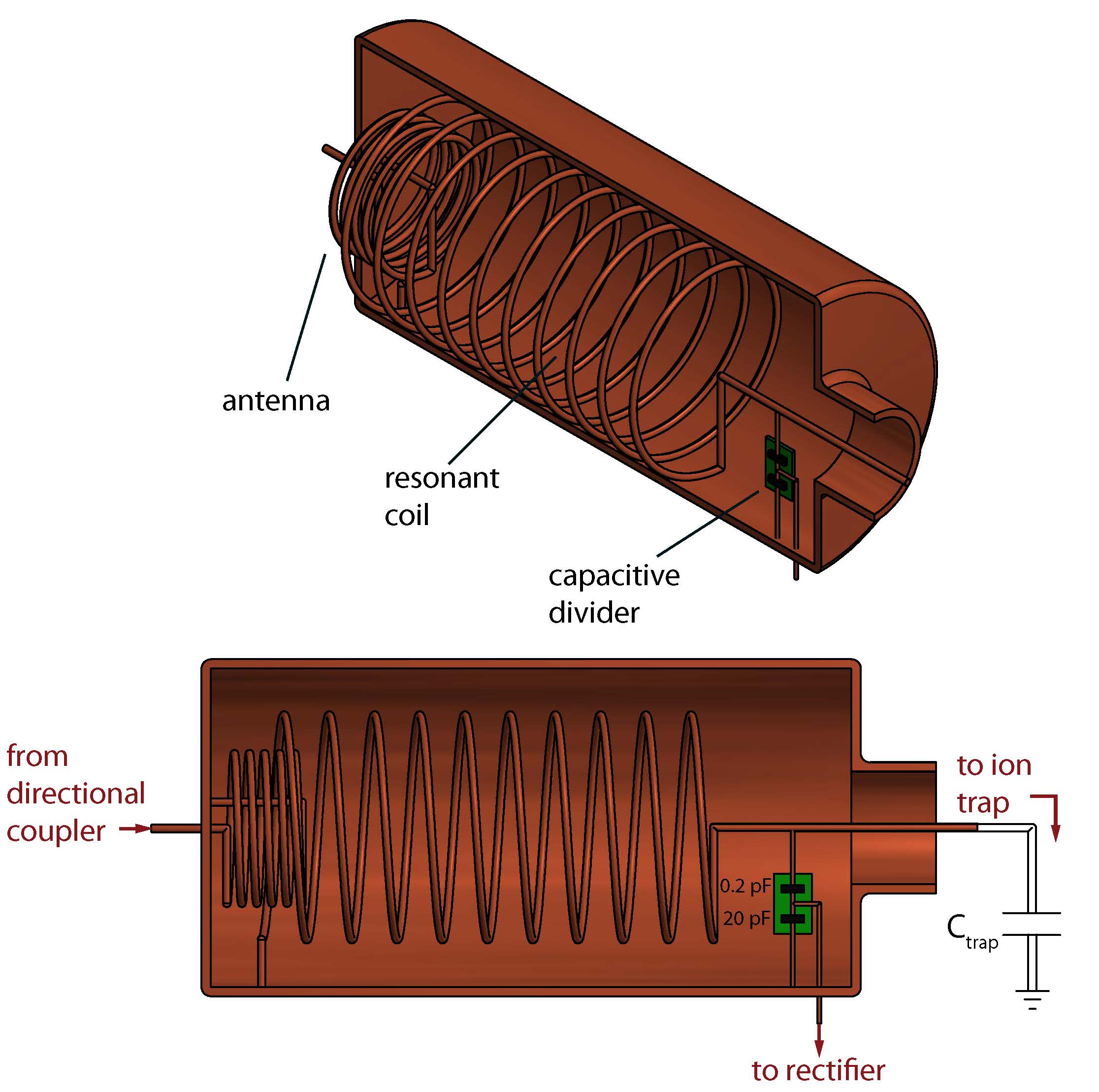}
	\caption{\label{pickoff_diagram.pdf} Helical quarter-wave resonator (transformer) with a 1:100 capacitive divider ($0.2$ pF and $20$ pF) mounted inside of the resonator near the high voltage side. The divider samples $V_0$ for feedback. A rigid wire is soldered from the output portion of the copper resonator coil to the copper-clad epoxy circuit board containing the dividing capacitors. The resonator drives the capacitance $C_{trap}$ of the vacuum feedthrough and ion trap electrodes.}
\end{figure}
\begin{figure*}
	\includegraphics[scale=.45]{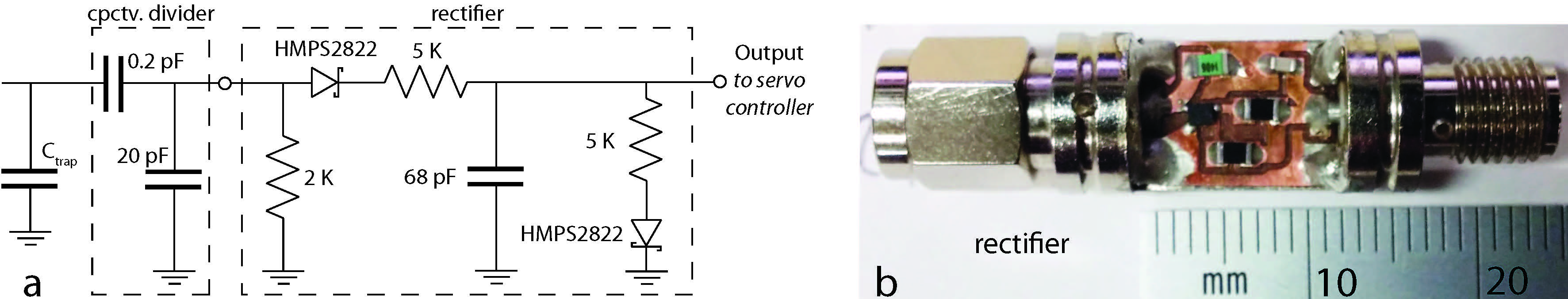}
	\caption{\label{rectifierandpickoff.pdf} (a) Schematic circuit diagram depicting the components of the pick-off voltage divider and temperature-compensating rectifier. (b) Photograph of the connectorized housing and mounted rectifier circuit.}
\end{figure*}

\section{trap rf stabilization}

We stabilize the rf confinement potential by sampling the high voltage rf signal supplying the ion trap electrode and feeding it back to a frequency mixer that controls the upstream rf oscillator amplitude.  As shown in the schematic of Fig. \ref{can_lock_schematic.pdf}b,  an rf signal at $\Omega/2\pi = 17$ MHz and $-8$dBm is produced by a function generator (SRS DS345) and sent through the Local Oscillator (LO) port of a level 3 frequency mixer (Mini-Circuits ZX05-1L-S), with a conversion loss of 5.6 dB. The rf port of the mixer is connected to a rf amplifier (Mini-Circuits TVA-R5-13) with a self-contained cooling system, providing a gain of 38 dB. The amplifier signal is fed into an antenna that inductively couples to a 17 MHz quarter-wave helical resonator and provides impedance matching between the rf source and the circuit formed by the resonator and ion trap electrode capacitance \cite{Siverns_12_APB}. The antenna, resonator, and equivalent ion trap capacitance $C_{trap}$ are shown in Fig. \ref{pickoff_diagram.pdf}, and exhibit an unloaded quality factor $Q_U\sim600$.

A capacitive divider samples roughly $1\%$ of the helical resonator output, using $C_1=0.2$ pF and $C_2=20$ pF ceramic capacitors (Vishay's QUAD HIFREQ Series) with temperature coefficients of \(0\pm30\) ppm$/^{\circ}$C.  With $C_1\ll C_{trap}$ and residual inductance between the divider and the trap electrodes much smaller than the resonator inductance itself, the divider faithfully samples the rf potential within a few centimeters of the trap electrodes and does not significantly load the trap/transformer circuit. The capacitors are surface-mounted to a milled copper-clad epoxy circuit board and installed inside the shielded resonator cavity, as diagrammed in Fig. \ref{pickoff_diagram.pdf}.

The sampled signal passes through a rectifier circuit (Fig. \ref{rectifierandpickoff.pdf}a) consisting of two Schottky diodes (Avago HMPS-2822 MiniPak) configured for passive temperature compensation \cite{Eriksson_00} and a low-pass filter giving a ripple amplitude 10 dB below the diode input signal amplitude. High quality foil resistors and ceramic capacitors are used to reduce the effect of temperature drifts. The entire rectifying circuit is mounted inside a brass housing (Crystek Corporation SMA-KIT-1.5MF) as shown in Fig. \ref{rectifierandpickoff.pdf}b. 
The sampling circuit has a bandwidth of $\sim 500$ kHz, limited by the 5 k$\Omega$/68 pF RC filter. 
The ratio of dc output voltage to rf input voltage amplitude, including the capacitive divider, is $1:250$ at a drive frequency of $17$ MHz, $1:330$ at $100$ MHz, and $1:870$ at a drive frequency of $1$ MHz.

\begin{figure}
	\includegraphics[scale=.45]{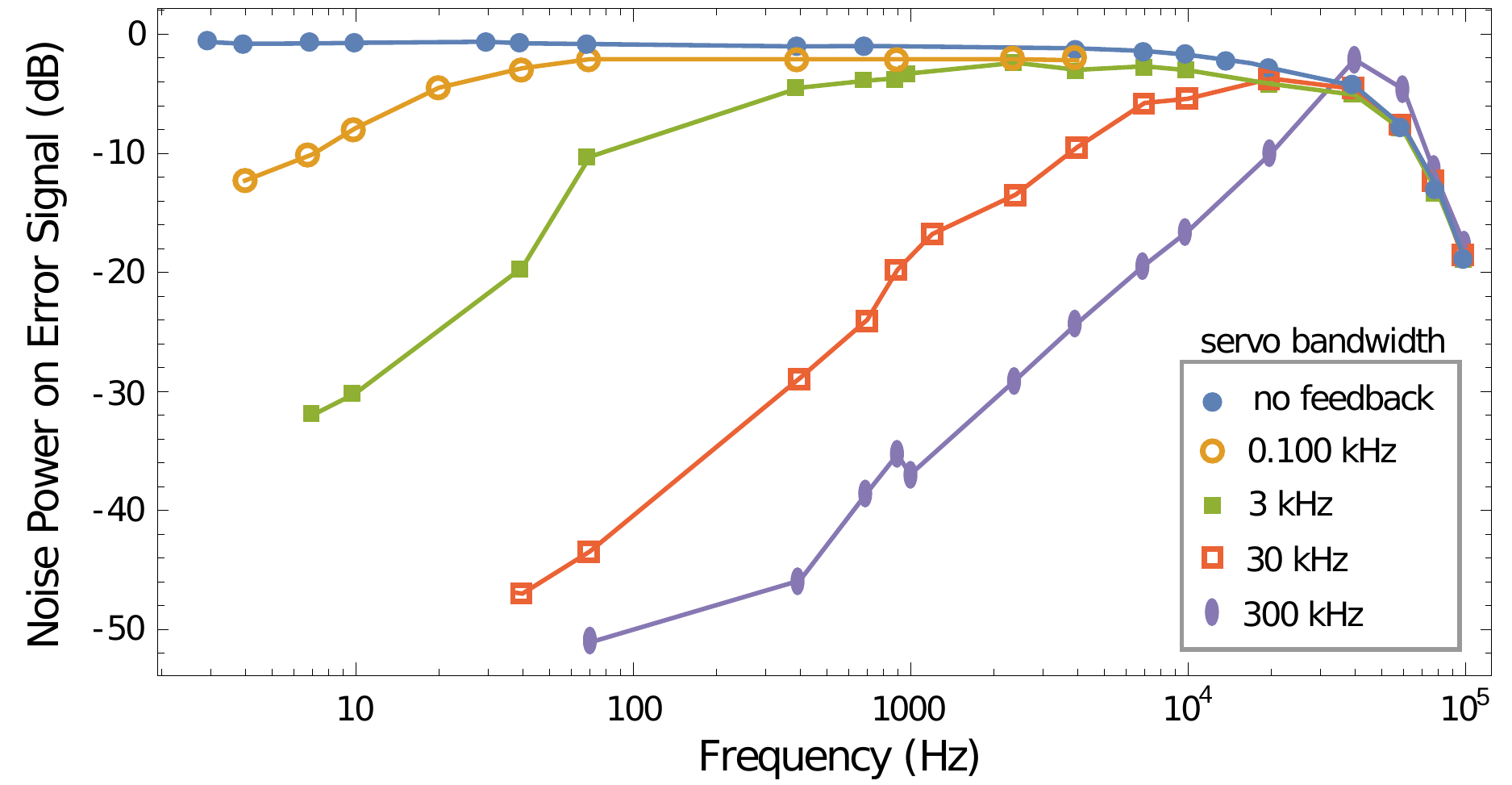}
	\caption{\label{transient.pdf} Suppression of injected noise in the stabilization circuit for various levels of feedback.  The rf drive is weakly amplitude-modulated at frequencies swept from 4 Hz to 100 kHz via a variable attenuator inserted before the rf amplifier. The amplitude of the resulting ripple on the error signal is measured as a function of the servo controller bandwidth.  The observed overall loop bandwidth of $\sim$30 kHz is consistent with the linewidth of the helical resonator transformer.}
\end{figure}

The dc rectified signal is compared to a stable set-point voltage (Linear Technology LTC6655 5V reference mounted on a DC2095A-C evaluation board) with variable control (Analog Devices EVAL-AD5791 and ADSP-BF527 interface board), giving 20-bit set-point precision and $\pm$0.25ppm stability.  The difference between these inputs -- the error signal -- is then amplified with proportional and integral gain (New Focus LB1005 servo controller) and fed back to regulate the upstream rf oscillator amplitude via the frequency mixer described above.  Figure \ref{transient.pdf} shows the response of the system for various servo controller gain settings when signals over a range of frequencies are injected into the system at the amplifier input.  The overall frequency response of the feedback loop is limited to a bandwidth of $30$ kHz, consistent with the linewidth $\Omega/(2\pi Q_U)$ of the helical resonator transformer.

\section{Ion Oscillation Frequency}

We next characterize the rf amplitude stabilization system by directly measuring the transverse motional oscillation frequency of a single atomic $^{171}$Yb$^+$ ion confined in the rf trap.  We perform optical Raman sideband spectroscopy \cite{Leibfried_03_RMP} on the $\ket{F=0,m_{f}=0} \equiv \ket{\downarrow}$ and $\ket{F=1,m_{f}=0}\equiv \ket{\uparrow}$ ``clock" hyperfine levels of the \(^{2}S_{1/2}\) electronic ground state of $^{171}$Yb$^+$.  This atomic transition has a frequency splitting of $\omega_{0}/2\pi=12.642815$ GHz and acquires frequency-modulated sidebands at $\omega_{0}\pm\omega$ due to the harmonic motion of the ion in the trap, with $\omega/2\pi \sim 1$ MHz. Before each measurement, the ion is Doppler cooled on the $^{2}S_{1/2}$ to $^{2}P_{1/2}$ electronic transition at a wavelength of $369.5$ nm \cite{Leibfried_03_RMP}. The ion is next prepared in the $\ket{\downarrow}$ state through optical pumping, and following the sideband spectroscopy described below, the state ($\ket{\downarrow}$ or $\ket{\uparrow}$) is measured with state-dependent fluorescence techniques \cite{Olmschenk_07_PRA}.

\begin{figure}
	\includegraphics[scale=.45]{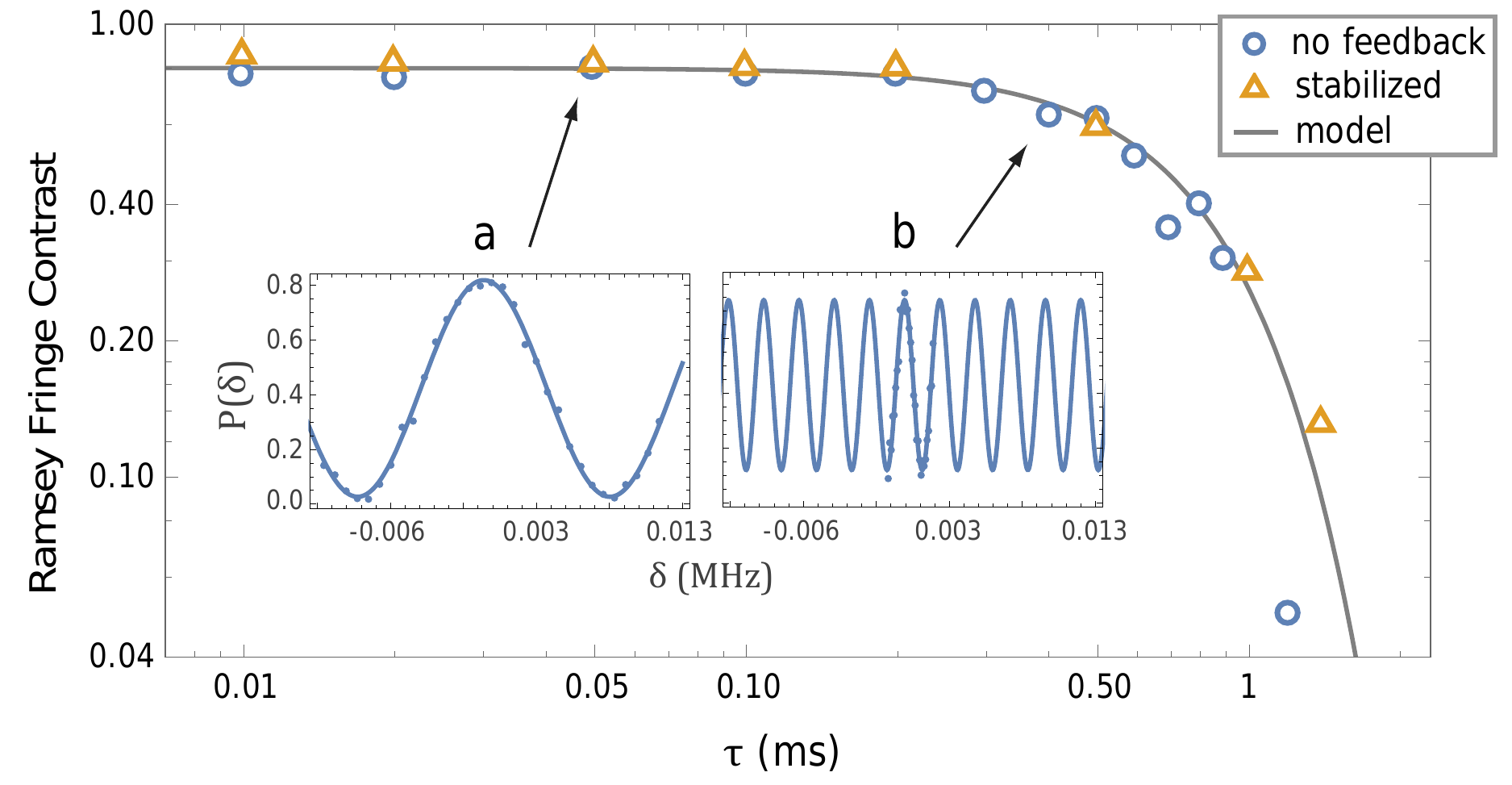}
	\caption{\label{noise_spectrum.pdf} Ramsey fringe contrast as a function of the Ramsey time $\tau$ between $\pi/2$ pulses, with and without feedback. The gray line is a model in which motional heating causes Ramsey fringe decoherence in $\sim 0.5$ ms. Inlays a and b show full Ramsey fringe measurements and fits for two different values of $\tau$.}
\end{figure}

The oscillation frequency is determined by performing Ramsey spectroscopy \cite{Ramsey_90_RMP} on the upper (blue) vibrational sideband of the clock transition at frequency $\omega_0+\omega$.  Because the atomic clock frequency $\omega_0$ is stable and accurate down to a level better than $1$ Hz, drifts and noise on the sideband frequency are dominated by the oscillation frequency $\omega$.  The sideband is driven by a stimulated Raman process from two counter-propagating laser light fields with a beatnote $\omega_L$ tuned near the upper vibrational sideband frequency \cite{Mizrahi_13_APB,Hayes_2010_PRL}. Following the usual Ramsey interferometric procedure \cite{Ramsey_90_RMP}, two $\pi/2$ pulses separated by time $\tau=0.4$ ms drive the Raman transition.  After the pulses are applied, the probability of finding the ion in the $\ket{\uparrow}$ state $P(\delta)=(1+C\cos\tau\delta)/2$ is sampled, where $\delta=\omega_L-(\omega_0+\omega)$ is the detuning of the beatnote from the sideband and $C$ is the contrast of the Ramsey fringes.  The Ramsey experiment is repeated 150 times for each value of $\delta$ in order to observe the Ramsey fringe pattern $P(\delta)$ and track the value of $\omega$.  Because this Raman transition involves a change in the motional quantum state of the ion, the Ramsey fringe contrast depends on the purity and coherence of atomic motion.  For short Ramsey times, the measured contrast of $\sim 0.8$ is limited by the initial thermal distribution of motional quantum states, and for Ramsey times $\tau>0.5$ ms, the fringe contrast degrades further (Fig. \ref{noise_spectrum.pdf}), which is consistent with a decoherence timescale $(2 \bar{n}_0 \dot{\bar{n}})^{-1}$ for initial thermal state $\bar{n}_0=15$ quanta and motional heating rate $\dot{\bar{n}}=100$ quanta/s \cite{Turchette_PRA_2000}.

\begin{figure}
	\includegraphics[scale=.45]{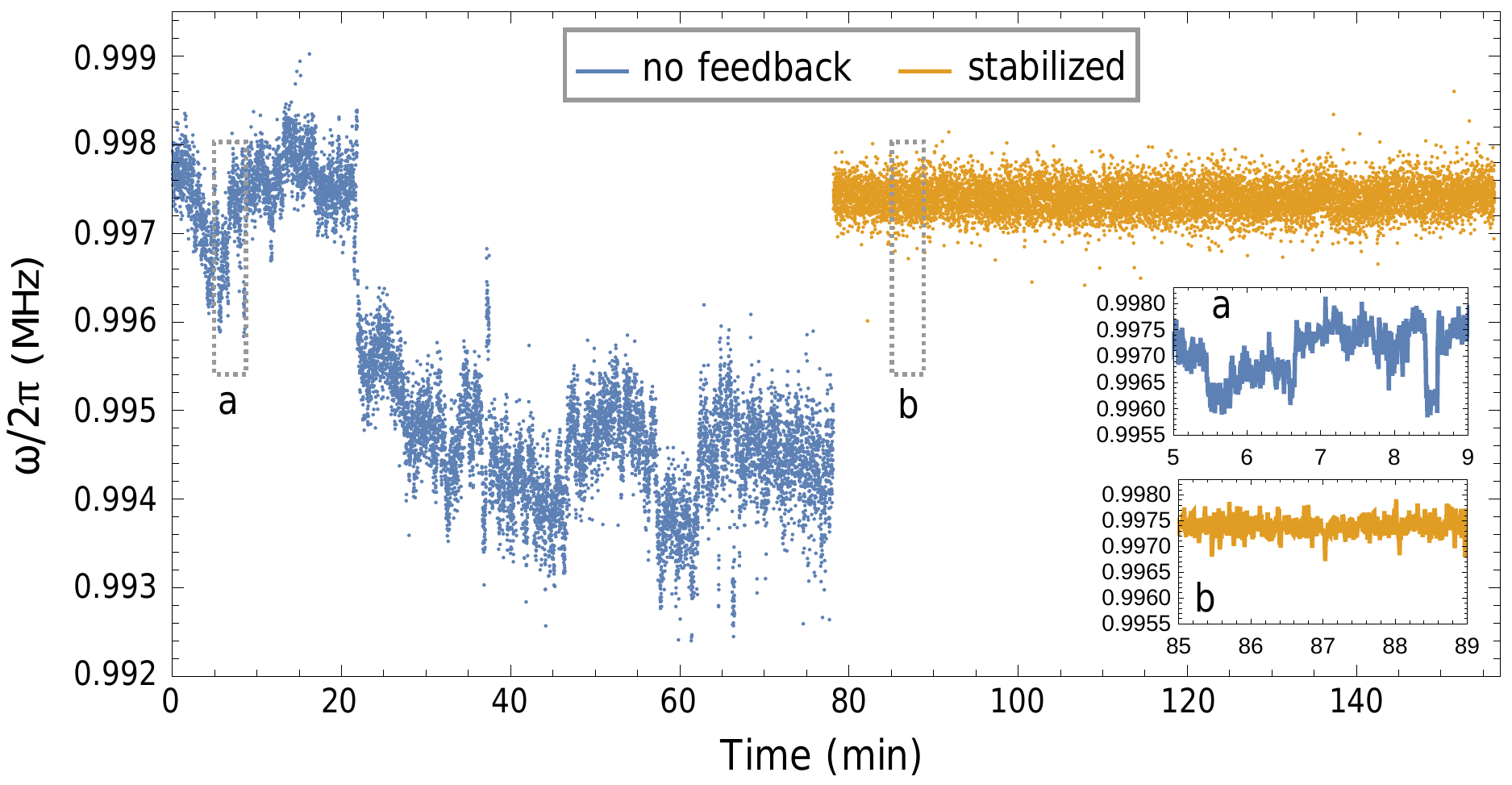}
	\caption{\label{long_time_drifts.pdf} Time dependence of the ion harmonic oscillation frequency $\omega$ plotted over the course of $160$ minutes with and without active stabilization.  With feedback there is a clear reduction in noise and drifts (apart from measurement shot noise, reflected by the fast fluctuations in the data).  Inlays a and b show magnified sections of the plot covering 4 minutes of integration.}
\end{figure}

Through Ramsey spectroscopy, we sample the ion trap oscillation frequency $\omega$ at a rate of 2.1 Hz for 80 minutes with no feedback on the rf potential, and then for another 80 minutes while actively stabilizing the rf potential. A typical time record of the the  measurements over these 160 minutes is shown in Fig. \ref{long_time_drifts.pdf}. Feedback control clearly improves the stability of the ion oscillation frequency, and we observe a $>30$ dB suppression of drifts over long times.  

\begin{figure}
	\includegraphics[scale=.45]{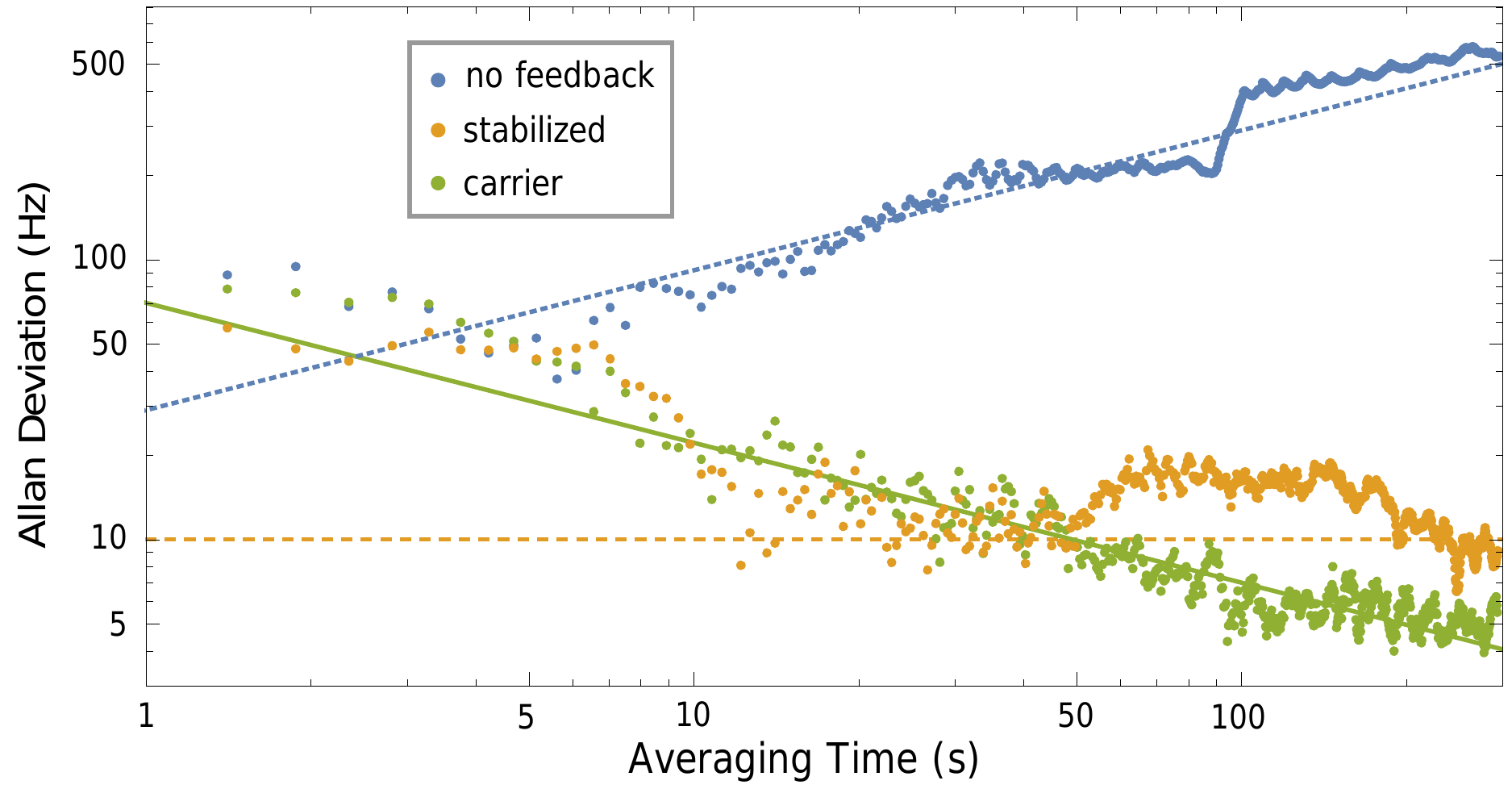}
	\caption{\label{allan_variance.pdf} Allan deviation data of the secular frequency $\omega$ while the system is with and without feedback, as well as the Allan deviation of the qubit carrier transition. The Allan deviation curves are calculated from the time record shown in Fig. \ref{long_time_drifts.pdf}, along with a similar measurement performed on the carrier transition.}
\end{figure}

From these measurements, we plot the Allan deviation \cite{Allan_66_IEEE} of the oscillation frequency in Fig. \ref{allan_variance.pdf} as a function of integration time $T$. When the system is stabilized, the Allan deviation in $\omega$ is nearly shot-noise limited (decreasing as $1/\sqrt{T}$) up to $\sim 200$ s of integration time, with a minimum uncertainty of better than 10 Hz, or 10 ppm, representing a $34$ dB suppression of ambient noise and drifts.  Without feedback, the trap frequency deviation drifts upward with integration time.  For integration times shorter than 7 s, there is not suffcient signal/noise in the measurements to see the effects of feedback stabilization.  However, as shown in Fig. \ref{transient.pdf}, the lock is able to respond to error signals up to a bandwidth of $\sim 30$ kHz, and we expect significant suppression of noise at these higher frequencies as well.  Although the Allan deviation of the oscillation frequency in the stabilized system improves with longer averaging time as expected, it drifts upward for a period just after $T=50$ s.  We confirm this drift appears in the ion oscillation frequency $\omega$ and not  the driving field $\omega_L$ or the ion hyperfine splitting $\omega_0$ by performing the same experiment on the clock ``carrier" transition near beatnote frequency $\omega_L = \omega_0$ instead of the upper sideband $\omega_L = \omega_0+\omega$.  As shown in Fig. \ref{allan_variance.pdf}, the measured Allan deviation of the carrier continues downward beyond $T=50$ s, meaning that the ion oscillation frequency is indeed the limiting factor at long times.

\section{Limits and noise sources}
It should be possible to stabilize the rf trap frequency much better than the observed 10 ppm by improving passive drifts outside of feedback control. These include the capacitive divider that samples the rf, the rectifier, the stable voltage reference, rf source frequency, and certain cables in the rf circuitry.  Most of these components will have residual drifts with temperature, mechanical strains, or other uncontrolled noise.  Below is a table of all crucial components outside of feedback control and their estimated contribution to the instability.
\\
\begin{center}
	\begin{tabular}{c c}
		\hline
		\\
		\textbf{Component} & \textbf{Stability}  \\
		\hline
		\\
		Capacitive Divider & $0 \leq 60$ ppm  \\
		Rectifier & $0.1$ ppm  \\
		Voltage Reference & $0.25$ ppm  \\
		rf source freq. & $0.1$ ppb  \\
		Cables & Unknown  \\
		\hline
	\end{tabular}
\end{center}

The main contribution is likely the capacitive divider, which is comprised of two capacitors each with a temperature coefficient of $\pm$ 30 ppm$/^{\circ}$C. Given the voltage divider configuration, the net temperature coefficient can range from $\sim 0-60$ ppm$/^{\circ}$C depending on how well the capacitors are matched.  Instabilities in the rectifier can arise from variability in the junction resistance of the diodes. In series with a 5 k$\Omega$ resistor, the $\sim 0.01 \Omega /^{\circ}$C junction resistance gives a net temperature coefficient of about $0.2$ ppm$/^{\circ}$C in the rectifier response. This is roughly equal to the temperature coefficient of the resistors used in the rectifier circuit. By using the temperature-stabilized circuit shown in Fig. \ref{rectifierandpickoff.pdf}, we estimate the net temperature coefficient of the rectifier response is reduced to $\sim$ 0.1 ppm$/^{\circ}$C.

Performance of the circuit is also helped by passively stabilizing components within the feedback loop as much as possible, such as temperature regulating the rf amplifier which feeds the resonator and using a passive mixer instead of a powered voltage variable attenuator. The helical transformer is particularly sensitive to temperature fluctuations and mechanical vibrations, which alters the resonance frequency and quality factor.  (Ensuring the helical coil is sealed against air currents can be more important than correcting small drifts in ambient temperature.)  If the resonant frequency of the transformer drifts too far, then a feedback circuit with a fixed frequency source (as used here and shown in Fig. \ref{can_lock_schematic.pdf}b) will call for more input power, and the servo system could possibly run away and become unstable. However, the resulting impedance mismatch from the off-resonant coupling will cause the servo to maintain the same amount of dissipated power in the resonator \cite{Siverns_12_APB} and not necessarily affect further drifts.  In any case, we do not observe such servo runaway.

\section{acknowledgments}
We thank N. M. Linke, S. Debnath, C. Figgatt, D. Hucul, P. W. Hess, C. Senko and K. Wright for useful discussions. This work is supported by the NSF Physics Frontier Center at JQI.

\bibliography{Kale_Johnson_Ultrafast}

\begin{thebibliography}{20}%
\makeatletter
\providecommand \@ifxundefined [1]{%
 \@ifx{#1\undefined}
}%
\providecommand \@ifnum [1]{%
 \ifnum #1\expandafter \@firstoftwo
 \else \expandafter \@secondoftwo
 \fi
}%
\providecommand \@ifx [1]{%
 \ifx #1\expandafter \@firstoftwo
 \else \expandafter \@secondoftwo
 \fi
}%
\providecommand \natexlab [1]{#1}%
\providecommand \enquote  [1]{``#1''}%
\providecommand \bibnamefont  [1]{#1}%
\providecommand \bibfnamefont [1]{#1}%
\providecommand \citenamefont [1]{#1}%
\providecommand \href@noop [0]{\@secondoftwo}%
\providecommand \href [0]{\begingroup \@sanitize@url \@href}%
\providecommand \@href[1]{\@@startlink{#1}\@@href}%
\providecommand \@@href[1]{\endgroup#1\@@endlink}%
\providecommand \@sanitize@url [0]{\catcode `\\12\catcode `\$12\catcode
  `\&12\catcode `\#12\catcode `\^12\catcode `\_12\catcode `\%12\relax}%
\providecommand \@@startlink[1]{}%
\providecommand \@@endlink[0]{}%
\providecommand \url  [0]{\begingroup\@sanitize@url \@url }%
\providecommand \@url [1]{\endgroup\@href {#1}{\urlprefix }}%
\providecommand \urlprefix  [0]{URL }%
\providecommand \Eprint [0]{\href }%
\providecommand \doibase [0]{http://dx.doi.org/}%
\providecommand \selectlanguage [0]{\@gobble}%
\providecommand \bibinfo  [0]{\@secondoftwo}%
\providecommand \bibfield  [0]{\@secondoftwo}%
\providecommand \translation [1]{[#1]}%
\providecommand \BibitemOpen [0]{}%
\providecommand \bibitemStop [0]{}%
\providecommand \bibitemNoStop [0]{.\EOS\space}%
\providecommand \EOS [0]{\spacefactor3000\relax}%
\providecommand \BibitemShut  [1]{\csname bibitem#1\endcsname}%
\let\auto@bib@innerbib\@empty
\bibitem [{\citenamefont {Dehmelt}(1990)}]{Dehmelt}%
  \BibitemOpen
  \bibfield  {author} {\bibinfo {author} {\bibfnamefont {H.}~\bibnamefont
  {Dehmelt}},\ }\href@noop {} {\bibfield  {journal} {\bibinfo  {journal} {Rev.
  Mod. Phys.}\ }\textbf {\bibinfo {volume} {62}},\ \bibinfo {pages} {525}
  (\bibinfo {year} {1990})}\BibitemShut {NoStop}%
\bibitem [{\citenamefont {Paul}(1990)}]{Paul}%
  \BibitemOpen
  \bibfield  {author} {\bibinfo {author} {\bibfnamefont {W.}~\bibnamefont
  {Paul}},\ }\href@noop {} {\bibfield  {journal} {\bibinfo  {journal} {Rev.
  Mod. Phys.}\ }\textbf {\bibinfo {volume} {62}},\ \bibinfo {pages} {531}
  (\bibinfo {year} {1990})}\BibitemShut {NoStop}%
\bibitem [{\citenamefont {Siverns}\ \emph {et~al.}(2012)\citenamefont
  {Siverns}, \citenamefont {Simkins}, \citenamefont {Weidt},\ and\
  \citenamefont {Hensinger}}]{Siverns_12_APB}%
  \BibitemOpen
  \bibfield  {author} {\bibinfo {author} {\bibfnamefont {J.~D.}\ \bibnamefont
  {Siverns}}, \bibinfo {author} {\bibfnamefont {L.~R.}\ \bibnamefont
  {Simkins}}, \bibinfo {author} {\bibfnamefont {S.}~\bibnamefont {Weidt}}, \
  and\ \bibinfo {author} {\bibfnamefont {W.~K.}\ \bibnamefont {Hensinger}},\
  }\href@noop {} {\bibfield  {journal} {\bibinfo  {journal} {Appl. Phys. B}\
  }\textbf {\bibinfo {volume} {107}},\ \bibinfo {pages} {921} (\bibinfo {year}
  {2012})}\BibitemShut {NoStop}%
\bibitem [{\citenamefont {Wineland}\ and\ \citenamefont
  {Blatt}(2008)}]{WinelandBlatt08}%
  \BibitemOpen
  \bibfield  {author} {\bibinfo {author} {\bibfnamefont {D.}~\bibnamefont
  {Wineland}}\ and\ \bibinfo {author} {\bibfnamefont {R.}~\bibnamefont
  {Blatt}},\ }\href@noop {} {\bibfield  {journal} {\bibinfo  {journal}
  {Nature}\ }\textbf {\bibinfo {volume} {453}},\ \bibinfo {pages} {1008}
  (\bibinfo {year} {2008})}\BibitemShut {NoStop}%
\bibitem [{\citenamefont {Monroe}\ and\ \citenamefont
  {Kim}(2013)}]{MonroeKimScience}%
  \BibitemOpen
  \bibfield  {author} {\bibinfo {author} {\bibfnamefont {C.}~\bibnamefont
  {Monroe}}\ and\ \bibinfo {author} {\bibfnamefont {J.}~\bibnamefont {Kim}},\
  }\href@noop {} {\bibfield  {journal} {\bibinfo  {journal} {Science}\ }\textbf
  {\bibinfo {volume} {339}},\ \bibinfo {pages} {1164} (\bibinfo {year}
  {2013})}\BibitemShut {NoStop}%
\bibitem [{\citenamefont {Richerme}\ \emph {et~al.}(2014)\citenamefont
  {Richerme}, \citenamefont {Gong}, \citenamefont {Lee}, \citenamefont {Senko},
  \citenamefont {Smith}, \citenamefont {Foss-Feig}, \citenamefont {Michalakis},
  \citenamefont {Gorshkov},\ and\ \citenamefont {Monroe}}]{Richerme2014}%
  \BibitemOpen
  \bibfield  {author} {\bibinfo {author} {\bibfnamefont {P.}~\bibnamefont
  {Richerme}}, \bibinfo {author} {\bibfnamefont {Z.-X.}\ \bibnamefont {Gong}},
  \bibinfo {author} {\bibfnamefont {A.}~\bibnamefont {Lee}}, \bibinfo {author}
  {\bibfnamefont {C.}~\bibnamefont {Senko}}, \bibinfo {author} {\bibfnamefont
  {J.}~\bibnamefont {Smith}}, \bibinfo {author} {\bibfnamefont
  {M.}~\bibnamefont {Foss-Feig}}, \bibinfo {author} {\bibfnamefont
  {S.}~\bibnamefont {Michalakis}}, \bibinfo {author} {\bibfnamefont {A.~V.}\
  \bibnamefont {Gorshkov}}, \ and\ \bibinfo {author} {\bibfnamefont
  {C.}~\bibnamefont {Monroe}},\ }\href@noop {} {\bibfield  {journal} {\bibinfo
  {journal} {Nature}\ }\textbf {\bibinfo {volume} {511}},\ \bibinfo {pages}
  {198} (\bibinfo {year} {2014})}\BibitemShut {NoStop}%
\bibitem [{\citenamefont {Jurcevic}\ \emph {et~al.}(2014)\citenamefont
  {Jurcevic}, \citenamefont {Lanyon}, \citenamefont {Hauke}, \citenamefont
  {Hempel}, \citenamefont {Zoller}, \citenamefont {Blatt},\ and\ \citenamefont
  {Roos}}]{Jurcevic2014}%
  \BibitemOpen
  \bibfield  {author} {\bibinfo {author} {\bibfnamefont {P.}~\bibnamefont
  {Jurcevic}}, \bibinfo {author} {\bibfnamefont {B.~P.}\ \bibnamefont
  {Lanyon}}, \bibinfo {author} {\bibfnamefont {P.}~\bibnamefont {Hauke}},
  \bibinfo {author} {\bibfnamefont {C.}~\bibnamefont {Hempel}}, \bibinfo
  {author} {\bibfnamefont {P.}~\bibnamefont {Zoller}}, \bibinfo {author}
  {\bibfnamefont {R.}~\bibnamefont {Blatt}}, \ and\ \bibinfo {author}
  {\bibfnamefont {C.~F.}\ \bibnamefont {Roos}},\ }\href@noop {} {\bibfield
  {journal} {\bibinfo  {journal} {Nature}\ }\textbf {\bibinfo {volume} {511}},\
  \bibinfo {pages} {202} (\bibinfo {year} {2014})}\BibitemShut {NoStop}%
\bibitem [{\citenamefont {Leibfried}\ \emph {et~al.}(2003)\citenamefont
  {Leibfried}, \citenamefont {Blatt}, \citenamefont {Monroe},\ and\
  \citenamefont {Wineland}}]{Leibfried_03_RMP}%
  \BibitemOpen
  \bibfield  {author} {\bibinfo {author} {\bibfnamefont {D.}~\bibnamefont
  {Leibfried}}, \bibinfo {author} {\bibfnamefont {R.}~\bibnamefont {Blatt}},
  \bibinfo {author} {\bibfnamefont {C.}~\bibnamefont {Monroe}}, \ and\ \bibinfo
  {author} {\bibfnamefont {D.}~\bibnamefont {Wineland}},\ }\href {\doibase
  10.1103/RevModPhys.75.281} {\bibfield  {journal} {\bibinfo  {journal} {Rev.
  Mod. Phys.}\ }\textbf {\bibinfo {volume} {75}},\ \bibinfo {pages} {281}
  (\bibinfo {year} {2003})}\BibitemShut {NoStop}%
\bibitem [{\citenamefont {Johnson}\ \emph {et~al.}(2015)\citenamefont
  {Johnson}, \citenamefont {Neyenhuis}, \citenamefont {Mizrahi}, \citenamefont
  {Wong-Campos},\ and\ \citenamefont {Monroe}}]{Johnson2015}%
  \BibitemOpen
  \bibfield  {author} {\bibinfo {author} {\bibfnamefont {K.~G.}\ \bibnamefont
  {Johnson}}, \bibinfo {author} {\bibfnamefont {B.}~\bibnamefont {Neyenhuis}},
  \bibinfo {author} {\bibfnamefont {J.}~\bibnamefont {Mizrahi}}, \bibinfo
  {author} {\bibfnamefont {J.~D.}\ \bibnamefont {Wong-Campos}}, \ and\ \bibinfo
  {author} {\bibfnamefont {C.}~\bibnamefont {Monroe}},\ }\href@noop {}
  {\bibfield  {journal} {\bibinfo  {journal} {Phys. Rev. Lett.}\ }\textbf
  {\bibinfo {volume} {115}},\ \bibinfo {pages} {213001} (\bibinfo {year}
  {2015})}\BibitemShut {NoStop}%
\bibitem [{\citenamefont {Chou}\ \emph {et~al.}(2010)\citenamefont {Chou},
  \citenamefont {Hume}, \citenamefont {Koelemeij}, \citenamefont {Wineland},\
  and\ \citenamefont {Rosenband}}]{NISTClock2010}%
  \BibitemOpen
  \bibfield  {author} {\bibinfo {author} {\bibfnamefont {C.~W.}\ \bibnamefont
  {Chou}}, \bibinfo {author} {\bibfnamefont {D.~B.}\ \bibnamefont {Hume}},
  \bibinfo {author} {\bibfnamefont {J.~C.~J.}\ \bibnamefont {Koelemeij}},
  \bibinfo {author} {\bibfnamefont {D.~J.}\ \bibnamefont {Wineland}}, \ and\
  \bibinfo {author} {\bibfnamefont {T.}~\bibnamefont {Rosenband}},\ }\href@noop
  {} {\bibfield  {journal} {\bibinfo  {journal} {Phys. Rev. Lett.}\ }\textbf
  {\bibinfo {volume} {104}},\ \bibinfo {pages} {070802} (\bibinfo {year}
  {2010})}\BibitemShut {NoStop}%
\bibitem [{\citenamefont {Raizen}\ \emph {et~al.}(1992)\citenamefont {Raizen},
  \citenamefont {Gilligan}, \citenamefont {Bergquist}, \citenamefont {Itano},\
  and\ \citenamefont {Wineland}}]{Raizen_1992_PRA}%
  \BibitemOpen
  \bibfield  {author} {\bibinfo {author} {\bibfnamefont {M.~G.}\ \bibnamefont
  {Raizen}}, \bibinfo {author} {\bibfnamefont {J.~M.}\ \bibnamefont
  {Gilligan}}, \bibinfo {author} {\bibfnamefont {J.~C.}\ \bibnamefont
  {Bergquist}}, \bibinfo {author} {\bibfnamefont {W.~M.}\ \bibnamefont
  {Itano}}, \ and\ \bibinfo {author} {\bibfnamefont {D.~J.}\ \bibnamefont
  {Wineland}},\ }\href@noop {} {\bibfield  {journal} {\bibinfo  {journal}
  {Phys. Rev. A}\ }\textbf {\bibinfo {volume} {45}},\ \bibinfo {pages} {6493}
  (\bibinfo {year} {1992})}\BibitemShut {NoStop}%
\bibitem [{\citenamefont {Zhu}, \citenamefont {Monroe},\ and\ \citenamefont
  {Duan}(2006)}]{Zhu_2006_PRL}%
  \BibitemOpen
  \bibfield  {author} {\bibinfo {author} {\bibfnamefont {S.-L.}\ \bibnamefont
  {Zhu}}, \bibinfo {author} {\bibfnamefont {C.}~\bibnamefont {Monroe}}, \ and\
  \bibinfo {author} {\bibfnamefont {L.-M.}\ \bibnamefont {Duan}},\ }\href@noop
  {} {\bibfield  {journal} {\bibinfo  {journal} {Phys. Rev. Lett.}\ }\textbf
  {\bibinfo {volume} {97}},\ \bibinfo {pages} {050505} (\bibinfo {year}
  {2006})}\BibitemShut {NoStop}%
\bibitem [{\citenamefont {Chiaverini}\ \emph {et~al.}(2005)\citenamefont
  {Chiaverini}, \citenamefont {Blakestad}, \citenamefont {Britton},
  \citenamefont {Jost}, \citenamefont {Langer}, \citenamefont {Leibfried},
  \citenamefont {Ozeri},\ and\ \citenamefont {Wineland}}]{Chiaverini_05_QIC}%
  \BibitemOpen
  \bibfield  {author} {\bibinfo {author} {\bibfnamefont {J.}~\bibnamefont
  {Chiaverini}}, \bibinfo {author} {\bibfnamefont {R.~B.}\ \bibnamefont
  {Blakestad}}, \bibinfo {author} {\bibfnamefont {J.}~\bibnamefont {Britton}},
  \bibinfo {author} {\bibfnamefont {J.~D.}\ \bibnamefont {Jost}}, \bibinfo
  {author} {\bibfnamefont {C.}~\bibnamefont {Langer}}, \bibinfo {author}
  {\bibfnamefont {D.}~\bibnamefont {Leibfried}}, \bibinfo {author}
  {\bibfnamefont {R.}~\bibnamefont {Ozeri}}, \ and\ \bibinfo {author}
  {\bibfnamefont {D.~J.}\ \bibnamefont {Wineland}},\ }\href@noop {} {\bibfield
  {journal} {\bibinfo  {journal} {Quantum Inf. Comput.}\ }\textbf {\bibinfo
  {volume} {5}},\ \bibinfo {pages} {419} (\bibinfo {year} {2005})}\BibitemShut
  {NoStop}%
\bibitem [{\citenamefont {Eriksson}\ and\ \citenamefont
  {Waugh}(2000)}]{Eriksson_00}%
  \BibitemOpen
  \bibfield  {author} {\bibinfo {author} {\bibfnamefont {H.}~\bibnamefont
  {Eriksson}}\ and\ \bibinfo {author} {\bibfnamefont {R.~W.}\ \bibnamefont
  {Waugh}},\ }\href@noop {} {\emph {\bibinfo {title} {A Temperature Compensated
  Linear Diode Detector, Design Tip}}},\ \bibinfo {organization} {Agilent
  Technologies} (\bibinfo {year} {2000})\BibitemShut {NoStop}%
\bibitem [{\citenamefont {Olmschenk}\ \emph {et~al.}(2007)\citenamefont
  {Olmschenk}, \citenamefont {Younge}, \citenamefont {Moehring}, \citenamefont
  {Matsukevich}, \citenamefont {Maunz},\ and\ \citenamefont
  {Monroe}}]{Olmschenk_07_PRA}%
  \BibitemOpen
  \bibfield  {author} {\bibinfo {author} {\bibfnamefont {S.}~\bibnamefont
  {Olmschenk}}, \bibinfo {author} {\bibfnamefont {K.~C.}\ \bibnamefont
  {Younge}}, \bibinfo {author} {\bibfnamefont {D.~L.}\ \bibnamefont
  {Moehring}}, \bibinfo {author} {\bibfnamefont {D.~N.}\ \bibnamefont
  {Matsukevich}}, \bibinfo {author} {\bibfnamefont {P.}~\bibnamefont {Maunz}},
  \ and\ \bibinfo {author} {\bibfnamefont {C.}~\bibnamefont {Monroe}},\
  }\href@noop {} {\bibfield  {journal} {\bibinfo  {journal} {Phys. Rev. A}\
  }\textbf {\bibinfo {volume} {76}},\ \bibinfo {pages} {052314} (\bibinfo
  {year} {2007})}\BibitemShut {NoStop}%
\bibitem [{\citenamefont {Ramsey}(1990)}]{Ramsey_90_RMP}%
  \BibitemOpen
  \bibfield  {author} {\bibinfo {author} {\bibfnamefont {N.~F.}\ \bibnamefont
  {Ramsey}},\ }\href@noop {} {\bibfield  {journal} {\bibinfo  {journal} {Rev.
  Mod. Phys.}\ }\textbf {\bibinfo {volume} {62}},\ \bibinfo {pages} {541}
  (\bibinfo {year} {1990})}\BibitemShut {NoStop}%
\bibitem [{\citenamefont {Mizrahi}\ \emph {et~al.}(2013)\citenamefont
  {Mizrahi}, \citenamefont {Neyenhuis}, \citenamefont {Johnson}, \citenamefont
  {Campbell}, \citenamefont {Senko}, \citenamefont {Hayes},\ and\ \citenamefont
  {Monroe}}]{Mizrahi_13_APB}%
  \BibitemOpen
  \bibfield  {author} {\bibinfo {author} {\bibfnamefont {J.}~\bibnamefont
  {Mizrahi}}, \bibinfo {author} {\bibnamefont {Neyenhuis}}, \bibinfo {author}
  {\bibfnamefont {K.~G.}\ \bibnamefont {Johnson}}, \bibinfo {author}
  {\bibfnamefont {W.~C.}\ \bibnamefont {Campbell}}, \bibinfo {author}
  {\bibfnamefont {C.}~\bibnamefont {Senko}}, \bibinfo {author} {\bibfnamefont
  {D.}~\bibnamefont {Hayes}}, \ and\ \bibinfo {author} {\bibfnamefont
  {C.}~\bibnamefont {Monroe}},\ }\href@noop {} {\bibfield  {journal} {\bibinfo
  {journal} {Appl. Phys. B}\ }\textbf {\bibinfo {volume} {114}},\ \bibinfo
  {pages} {45} (\bibinfo {year} {2013})}\BibitemShut {NoStop}%
\bibitem [{\citenamefont {Hayes}\ \emph {et~al.}(2010)\citenamefont {Hayes},
  \citenamefont {Matsukevich}, \citenamefont {Maunz}, \citenamefont {Hucul},
  \citenamefont {Quraishi}, \citenamefont {Olmschenk}, \citenamefont
  {Campbell}, \citenamefont {Mizrahi}, \citenamefont {Senko},\ and\
  \citenamefont {Monroe}}]{Hayes_2010_PRL}%
  \BibitemOpen
  \bibfield  {author} {\bibinfo {author} {\bibfnamefont {D.}~\bibnamefont
  {Hayes}}, \bibinfo {author} {\bibfnamefont {D.~N.}\ \bibnamefont
  {Matsukevich}}, \bibinfo {author} {\bibfnamefont {P.}~\bibnamefont {Maunz}},
  \bibinfo {author} {\bibfnamefont {D.}~\bibnamefont {Hucul}}, \bibinfo
  {author} {\bibfnamefont {Q.}~\bibnamefont {Quraishi}}, \bibinfo {author}
  {\bibfnamefont {S.}~\bibnamefont {Olmschenk}}, \bibinfo {author}
  {\bibfnamefont {W.}~\bibnamefont {Campbell}}, \bibinfo {author}
  {\bibfnamefont {J.}~\bibnamefont {Mizrahi}}, \bibinfo {author} {\bibfnamefont
  {C.}~\bibnamefont {Senko}}, \ and\ \bibinfo {author} {\bibfnamefont
  {C.}~\bibnamefont {Monroe}},\ }\href@noop {} {\bibfield  {journal} {\bibinfo
  {journal} {Phys. Rev. Lett.}\ }\textbf {\bibinfo {volume} {104}},\ \bibinfo
  {pages} {140501} (\bibinfo {year} {2010})}\BibitemShut {NoStop}%
\bibitem [{\citenamefont {Turchette}\ \emph {et~al.}(2000)\citenamefont
  {Turchette}, \citenamefont {Myatt}, \citenamefont {King}, \citenamefont
  {Sackett}, \citenamefont {Kielpinski}, \citenamefont {Itano}, \citenamefont
  {Monroe},\ and\ \citenamefont {Wineland}}]{Turchette_PRA_2000}%
  \BibitemOpen
  \bibfield  {author} {\bibinfo {author} {\bibfnamefont {Q.~A.}\ \bibnamefont
  {Turchette}}, \bibinfo {author} {\bibfnamefont {C.~J.}\ \bibnamefont
  {Myatt}}, \bibinfo {author} {\bibfnamefont {B.~E.}\ \bibnamefont {King}},
  \bibinfo {author} {\bibfnamefont {C.~A.}\ \bibnamefont {Sackett}}, \bibinfo
  {author} {\bibfnamefont {D.}~\bibnamefont {Kielpinski}}, \bibinfo {author}
  {\bibfnamefont {W.~M.}\ \bibnamefont {Itano}}, \bibinfo {author}
  {\bibfnamefont {C.}~\bibnamefont {Monroe}}, \ and\ \bibinfo {author}
  {\bibfnamefont {D.~J.}\ \bibnamefont {Wineland}},\ }\href@noop {} {\bibfield
  {journal} {\bibinfo  {journal} {Phys. Rev. A}\ }\textbf {\bibinfo {volume}
  {62}},\ \bibinfo {pages} {053807} (\bibinfo {year} {2000})}\BibitemShut
  {NoStop}%
\bibitem [{\citenamefont {Allan}(1966)}]{Allan_66_IEEE}%
  \BibitemOpen
  \bibfield  {author} {\bibinfo {author} {\bibfnamefont {D.~W.}\ \bibnamefont
  {Allan}},\ }in\ \href@noop {} {\emph {\bibinfo {booktitle} {Proc. IEEE}}},\
  Vol.~\bibinfo {volume} {54}\ (\bibinfo {year} {1966})\ p.\ \bibinfo {pages}
  {221}\BibitemShut {NoStop}%
\end{thebibliography}%
\end{document}